\documentclass{aa}
\usepackage{graphics,epsfig}
\usepackage{amssymb}

\newcommand{\kms}{km~s$^{-1}$}

\begin{document}

\title{Kinematics of two dwarf galaxies in the NGC~6946 group}
\titlerunning{Kinematics of two dwarf galaxies}
\author{Ayesha Begum\inst{1}\thanks{ayesha@ncra.tifr.res.in} and
        Jayaram N. Chengalur\inst{1}\thanks{chengalur@ncra.tifr.res.in}
}
\authorrunning{Begum \& Chengalur}
\institute{National Centre for Radio Astrophysics,
	Post Bag 3, Ganeshkhind, Pune 411 007}
\date{Received mmddyy/ accepted mmddyy}
\offprints{Ayesha Begum}
\abstract{
We present high velocity resolution ($\sim 1.6$~\kms)
Giant Meterwave Radio Telescope (GMRT) HI 21 cm synthesis images of 
the dwarf galaxies KK98 250 and KK98 251, as well as optical broad band images
of KK98 250. We find that, despite being faint ($\rm{M_B} \lesssim -14$), both
galaxies have regular velocity fields. In both cases the
velocity fields are consistent with rigid body rotation. We derive 
rotation curves for the galaxies and fit mass models to them. For both
galaxies, we find acceptable fits using  isothermal halos while NFW halos 
provide a poor fit.  Finally, we compile from literature a sample of
galaxies with HI synthesis observations and I band magnitudes. From 
this sample we  find, in agreement with earlier studies (which used 
single dish HI data),  that dwarf galaxies tend to lie below the 
I band Tully-Fisher relation defined by brighter galaxies.
}
\maketitle

\section{Introduction}
\label{intro}

There has been renewed interest in the kinematics of dwarf galaxies for two reasons. 
(1)~There appears to be a conflict between hierarchical galaxy formation model
predictions for, and observational determinations of, the dark matter halo density 
profile of galaxies. In particular, these models predict a cuspy central halo, 
which many observers find inconsistent with observations (e.g. ~\cite{weldrake03}). 
While the models predict
cuspy halos for all galaxies, dwarf galaxies are believed to be best suited for 
tests of the models because these galaxies are in general held to be
dominated by dark matter, even in the innermost regions. In larger galaxies, uncertainties
in the stellar mass to light ratio limit one's ability to determine density profile  of 
the dark matter halo. (2)~It appears that faint dwarf galaxies deviate significantly
from the Tully-Fisher relationship as defined by bright galaxies. While the TF relation
is generally studied using the inclination corrected 50\% profile width (usually 
obtained from single dish observations), it is unclear whether this is an appropriate
measure to use for the faintest dwarf galaxies, where the rotational velocities 
are comparable to the velocity dispersion. For such galaxies, the ``asymmetric drift''
corrected rotational velocity is probably a better measure. However, estimating
this quantity requires HI synthesis imaging. Further, for the faint dwarf irregulars,
even the inclination may be difficult to estimate from the optical image and may
only be obtainable from the HI distribution and kinematics. Unfortunately, the
number of faint dwarf galaxies with sufficiently high quality HI synthesis images
available is quite small. We discuss here GMRT HI observations of two dwarf 
galaxies, KK98 250 and KK98 251, in light of both these issues. In discussing the TF 
relation, we  combine our data for KK98 250 
and KK98 251 with our earlier data for still fainter galaxies and data for brighter 
galaxies available from the literature. 

     Both KK98 250 and KK98 251 were identified as companions to the giant 
spiral galaxy NGC 6946 by~\cite{karachentseva98}, and 
have also been observed in HI at Effelsberg 
(Huchtmeier et al. 1997) and at the DRAO (\cite{pisano00}). The data presented 
here are of much higher sensitivity and resolution than the DRAO data. 
Distance estimates to KK98 250 and KK98 251 vary from 5.3 Mpc (Sharina et al. 1997, 
Karachentsev et al. 2000) to 8.2~Mpc (Sharina et al. 1997). Because of the proximity of 
the two galaxies to each other and to NGC 6946, on the sky as well as in the velocity,
we feel that it is likely that all these galaxies belong to the same group viz. the
NGC 6946 group. Hence, in this paper we take the mean distance to the group (5.6~Mpc, 
estimated from the brightest stars in eight members of the group) as the  distance 
to both the galaxies (Huchtmeier et al. 2000). At this distance, the absolute blue magnitude for KK98 250 and 
KK98 251 are $\rm{M_B}\sim -$13.72 and $-$14.54 respectively.

The rest of the paper is divided as follows. The GMRT observations are detailed 
in Sect.~\ref{sec:obs}, while the results 
are presented and discussed in Sect.~\ref{sec:res}. 

\section{Observations}
\label{sec:obs}     

\subsection{Optical Observations and analysis}
\label{ssec:opt_obs}

        CCD images of KK98 250 and KK98 251 in the Bessell~I and V filters were obtained 
on 29$^{\rm{th}}$ and 30$^{\rm{th}}$ May 2003, using the HFOSC (Himalayan Faint Object Spectrograph 
Camera), at the 2.0 m Himalayan Chandra Telescope. The camera has a field of view of 10$'\times$10$'$,
with a scale of 0.3$^{''}$/pixel. Both the galaxies were covered in the same pointing. 
The total exposure time on the target was 40 min in I and 50 min in the V band. The FWHM 
seeing of the co-added images was $\sim1.7^{''}$.  Since the first night was not 
photometric, standard fields (from ~\cite{landolt83}) were observed only on the 
second night. 

         Debiasing, flat-fielding and cosmic ray filtering were done in the 
usual manner, using standard IRAF routines. The exposures taken on the 30$^{\rm{th}}$ were 
calibrated and then added after alignment. This combined frame was used to calibrate 
the individual exposures taken on the 29$^{\rm{th}}$. No fringe subtraction was done for the 
I band images. 

       Surface photometry for KK98 250 was done using the ellipse fitting algorithm 
of~\cite{bender87}. Ellipse fitting in the I band was restricted to 
the inner  regions of the galaxy, since the outer low surface brightness emission
is affected by fringing.  The average ellipticity was found to be $0.72\pm0.02$ 
(corresponding to an inclination of $\sim 79^\circ$, for an intrinsic thickness 
ratio $q_0 =0.2$). The position angle, measured between the north direction
on the sky and the major axis of the receding half of the galaxy (see 
Sect.~\ref{ssec:rotcur}), was found to be $\sim 270^\circ$.
The surface brightness profiles obtained from the ellipse fitting in 
V and I bands  are well described by exponential disk with the scale 
length  of ${41.0''}$ ($\sim$1.1 kpc) and ${62.0''}$ ($\sim$1.7 kpc) 
respectively. The average $<$V-I$>$ color is $\sim$1.2, with 
little variation with radius. 
The total magnitude after correcting for galactic extinction (using $A_V$=1.01
~mag and $A_I$=0.59~mag; Schlegel et al. 1998) is  13.8$\pm$0.3 mag in V and 
12.6$\pm$0.4 mag in I. No correction for internal extinction was applied. 

     KK98 251 has several bright stars superposed on it which could not be accurately 
removed. For this galaxy, we hence use the results of the I band photometry derived 
by~\cite{karachentesv00}, who derived an I band exponential scale length  and 
a total I magnitude of ${22.8''}$ ($\sim$0.6 kpc) and 14.42 mag respectively.

\subsection{HI observations and analysis}
\label{ssec:gmrt_obs}

\begin{table}[b!]
\caption{Parameters of the GMRT observations}
\label{tab:obs}
\vskip 0.1in
\begin{tabular}{ll}
\hline
Parameters& Value \\
\hline
\hline
RA(2000) & 20$^h$30$^m$33.0$^s$\\
Declination(2000) &  ${60}^{\circ} 21' 17''$\\
Central velocity (heliocentric) & 121.0~\kms\\
Date of observations & 22 June 2001\\
Time on source & 6 hrs\\
Total bandwidth & 1.0 MHz\\
Number of channels & 128\\
Channel separation & 1.65~\kms\\
FWHM of synthesized beam  & 43$''\times38''$, 26$''\times21''$,\\ 
                          & 16$''\times14''$, 11.5$''\times10''$ \\
RMS noise per channel & 3.5~mJy, 2.8 mJy, 2.2 mJy\\ 
                           &1.9 mJy\\
\hline
\end{tabular}
\end{table}
    
The GMRT (Swarup et al. 1991) observations of KK98 250 and KK98 251 were conducted during 
the commissioning phase of the telescope. The field of view of the telescope 
($\sim 24^{'})$ is large enough to cover both galaxies in a single pointing. 
The setup for the  observations is given  in Table~\ref{tab:obs}. The calibrators
were 3C48 and  3C286 (flux), 2022+616 (phase) and 3C286 (bandpass).

     The data were reduced in the usual way using standard tasks in classic AIPS. 
The GMRT  has a hybrid configuration which simultaneously provides both high angular 
resolution ($\sim 2^{''}$ if one uses baselines between the arm antennas) as well as
sensitivity to extended emission (from baselines between the antennas in the 
central array). Data cubes were therefore made using various (u,v) cutoffs, 
including 0$-$5 k$\lambda$, 0$-$10 k$\lambda$, 0$-15$ k$\lambda$ and 0$-$20 k$\lambda$.
The corresponding angular resolutions are 43$''\times38''$, 26$''\times21''$, 
16$''\times14''$ and  11.5$''\times10''$. All the data cubes were deconvolved
using the AIPS task IMAGR and then corrected for primary beam attenuation.

     Moment maps  of the data cubes were made using the AIPS task MOMNT. Maps of the
velocity field and line profile widths were also made in GIPSY, using  single gaussian 
fits to the line profiles. The velocity fields produced by gaussian fitting were 
found to be in a reasonable agreement with that obtained from moment analysis. 
As we discuss in more detail later, we need to know the velocity dispersion 
of the gas in order to estimate the pressure support in the HI disk. For dwarf
galaxies, where the contribution from systematic rotation to the profile width 
in a single synthesized beam is small, the velocity dispersion can generally be 
estimated from the width of the line profiles. However, for a highly inclined
galaxy like KK98 250, the profile width is dominated by rotational motion even within
a single synthesized beam. For this galaxy, we hence assume that the velocity 
dispersion is $\sim 8$~\kms, a typical value for dwarf galaxies 
(e.g. \cite{lake90,skillman88}). For KK98 251, from the widths obtained from 
gaussian fitting (which we prefer to use because the thresholding algorithm 
used by the MOMNT task results in an underestimation of the true profile width), 
we find that the profile widths vary from $\sim 8.4 \pm 0.7$~\kms~in 
the inner region (upto $80^{''}$) to $\sim 9.5 \pm 0.9$~\kms in the outer regions.

     Finally, a continuum image was made using the average of the line free channels 
but no extended continuum (to  $3\sigma$ limit of 1.8~mJy/beam for a beam size 
of $26^{''}\times22^{''}$) was detected from either galaxy. A high
resolution ($5.8^{''}\times5.2^{''}$) image also showed no compact sources
associated with either galaxy down to a $3\sigma$ limit of 1.0~mJy/beam.

\section{Results and Discussion}
\label{sec:res}
\subsection{HI distribution}
\label{ssec:HI_dis}

\begin{figure}[t!]
\epsfig{file=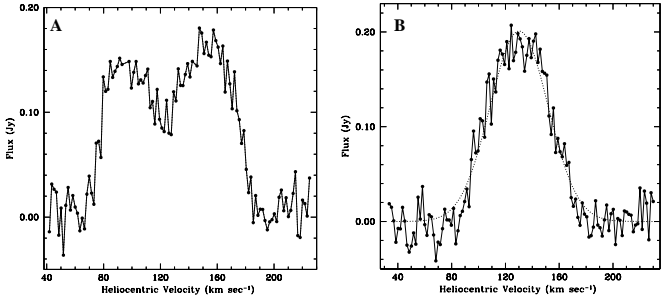,width=3.6in}
\caption{ 
          {\bf[A]}HI profile for KK98 250 obtained from 43$^{''}\times38^{''}$ data cube.
          The channel separation is $1.65$~\kms. Integration of profile 
          gives a flux integral of  $16.4$~Jy \kms and an HI mass of
          $12.1\times{10}^{7} \rm{M_\odot}$. 
{ \bf[B]} HI profile for KK98 251 obtained from 43$^{''}\times38^{''}$ data cube.
          The channel separation is $1.65$~\kms. Integration of profile
          gives a flux integral of  $10.6$~Jy \kms and an HI mass of
          $7.8\times{10}^{7} \rm{M_\odot}$. The dotted line shows
          a gaussian fit to the profile.
         }
\label{fig:HI_spec}
\end{figure}

   The global HI emission profiles of KK98 250 and KK98 251 (obtained from 43$''\times38''$
data cube), are shown in Fig.~\ref{fig:HI_spec}. 
The integrated flux of KK98 250 is $16.4\pm 1.6$~Jy~\kms, the (heliocentric) systemic 
velocity and the velocity width at 50\% level of peak emission (W$_{50}$) are 
126.0$\pm 1.6$~\kms and $95.5\pm2.3$~\kms. 
The systemic velocity is a good match to the single dish value of $127.0 \pm 2.0$~\kms 
(\cite{hucht97}), but the flux  integral is considerably lower than the single dish 
value of 20.0~Jy~\kms. For KK98 251, the integrated flux is $10.6\pm 1.0$~Jy~\kms, the 
systemic velocity is  $130.2\pm1.7$~\kms and W$_{50}$ is $51.7\pm1.8$~\kms. 
Again, the systemic velocity is in reasonable agreement with the value of 
$126.0 \pm2.0$~\kms obtained from single dish observations (\cite{hucht97}), but
the integrated flux and W$_{50}$ are considerably smaller than the single 
dish values of 14.6~Jy~\kms and $64.0$~\kms respectively. 

      The GMRT fluxes could be lower than those obtained from single dish measurements 
either because of (i)~a calibration error or (ii)~a large fraction of the HI being in 
an extended distribution that is resolved out.  However, the flux of the point sources 
seen in the GMRT image are in good agreement with those listed in NVSS, indicating that
our calibration is not at fault. Further, from our past experience in HI imaging 
of galaxies (with sizes similar to  KK98 250 and KK98 251) with the GMRT, it seems unlikely 
that we have resolved out a large fraction of the total flux. Interestingly, a large 
discrepancy between the interferometric fluxes and the single dish fluxes was also 
seen in  the DRAO images of these galaxies (\cite{pisano00}), although the comparison 
in that case is complicated by the very poor signal to noise ratio of the DRAO data. 
Finally, we note that for both the galaxies there is a strong local HI emission 
at velocities very close to the systemic velocities. Hence, it is likely that the single 
dish  integrated flux measurements were contaminated by  blending of the HI emission 
from the galaxies with that of the  galactic emission due to both the coarse 
velocity resolution ($\sim$10.0 \kms), as well as imperfect subtraction of the
foreground emission in the position switching mode used
in those observations. If we assume that the 
total fluxes are those measured at the GMRT, then the HI mass of  KK98 250 is 
$12.1\pm1.1 \times{10}^{7}~\rm{M_\odot}$ and $\rm{M_{\rm{HI}}/L}_{\rm{B}}\sim 1.2$.
For KK98 251, the corresponding numbers are $7.8\pm0.7 \times{10}^{7}~\rm{M_\odot}$ 
and $1.6$.

\begin{figure}[h]
\epsfig{file=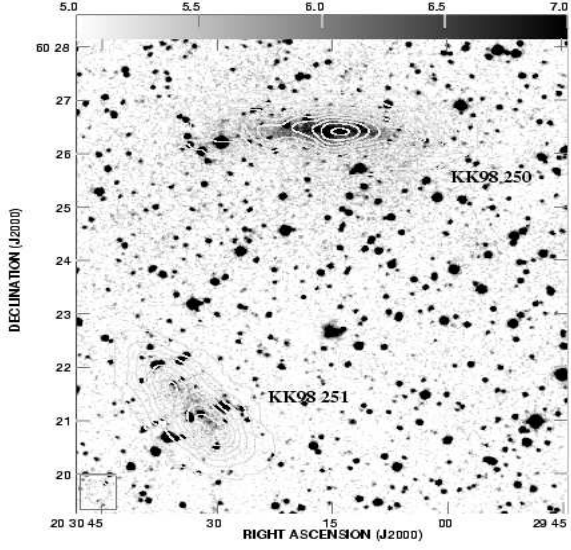,width=4.0in,height=3.0in}
\caption{The optical DSS image of KK98 250 (north) and KK98 251 (south-east), shown 
         in greyscales, with the GMRT 26$^{''}\times21^{''}$  resolution integrated HI
         emission map (contours) overlayed. The contour levels are 0.02, 2.8, 5.7, 8.7,
11.6, 14.4, 17.5, 20.1, 22.9, 25.8, 28.9, 31.7, 34.4, 37.4 and 40.2 $\times$10$^{20}$ 
atoms cm$^{-2}$.}
\label{fig:ov}
\end{figure}

Fig.~\ref{fig:ov} shows the integrated HI emission from KK98 250 and KK98 251  at
26$''\times21''$ resolution, overlayed on the digitized sky survey (DSS)
image. Although KK98 250 may be mildly warped, there is no clear signature 
of interaction between the two galaxies. For KK98 251, two high density peaks
are seen near the center.

An estimate of the morphological center, position angle (PA) and inclination 
(assuming an intrinsic thickness ratio $q_0 = 0.2$) of the galaxies were obtained
by fitting elliptical annuli to the 43$''\times38''$, 26$''\times21''$  and 
16$''\times14''$ resolution integrated HI column density maps. For KK98 250, the  inclination and 
PA were  found to be $80\pm 4$ and $267\pm3$ degrees respectively. Due to the high
density clumps in the center, ellipse fitting for KK98 251 was restricted to the 
outer contours of the 43$''\times38''$ and 26$''\times21''$ resolution images.
The inclination and PA of the HI  disk were found to be  $62\pm 5$ and
$220\pm5$ degrees respectively. For each galaxy, the value of the
PA and inclination  estimated from different resolution images  match
within the  error bars.  For KK98 250, the estimated parameters are also
in good agreement with the values obtained from the optical image. 

\begin{figure}[h]
\epsfig{file=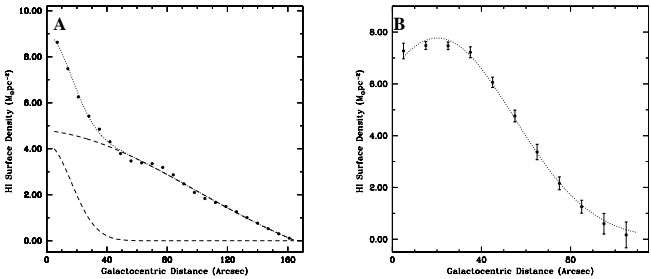,width=3.5in}
\caption{ {\bf [A]} The HI surface density profile of KK98 250 derived from the HI
          distribution at $26^{''}\times23^{''}$  resolution (points). A double 
          gaussian fit to the HI distribution is shown as dotted line. A decomposition 
         of the double gaussian fit into a narrow and broad gaussian component is 
         shown as   dashed lines. See the text for more details.  
	  {\bf [B]} The HI surface density profile of KK98 251 derived from the HI
          distribution at $26^{''}\times23^{''}$  resolution (points). The adopted 
          gaussian fit is shown as a dotted line. 
        }
\label{fig:smd}
\end{figure}

Another input that we require for estimating the pressure support of the HI
disk is the deprojected HI radial surface density profile. For highly inclined galaxies,
deprojection using ellipse fitting does not lead to reliable estimates for the surface
density. Hence, in the case of KK98 250, Lucy's (1974) iterative decovolution scheme,
as adapted and developed by~\cite{warmel88}, was used to derive the HI surface mass
density. Fig.~\ref{fig:smd}[A] shows the best fit HI radial profile obtained from the 
26$''\times21''$  resolution HI image using Lucy's scheme. Note that the deconvolved 
HI surface density profile shows a  steep rise of  $\sim$40\% in the central 
$\sim 25^{''}$, i.e. within one synthesized beam. This is likely to be an artifact 
produced by the deconvolution; similar artifacts produced by this method have 
been seen earlier, by for e.g.~Swaters~(1999). We therefore fit the surface 
density $\Sigma_{\rm{HI}}(r)$ by a double Gaussian (shown as dotted line in 
Fig.~\ref{fig:smd}[A]), and use only the broad component for further analysis. 
For reference the full double Gaussian fit is given by:

\begin{equation}
\Sigma_{\rm{HI}}(r)=\Sigma_{01}\times e^{-r^2/2r^2_{01}}+\Sigma_{02}\times e^{-r^2/2r^2_{02}}
\label{eqn:hisb2}
\end{equation}

with $r_{01}=73.9^{''}\pm1.7^{''}$, $r_{02}=14.4^{''}\pm0.8^{''}$, $\Sigma_{01}=
5.0\pm0.1 \rm{M_\odot}~pc^{-2}$ and $\Sigma_{02}=4.2\pm0.2 \rm{M_\odot}~pc^{-2}$.

For KK98 251, the HI surface density profile was obtained in the usual way from
fitting elliptical annuli. The surface density  $\Sigma_{\rm{HI}}(r)$ (shown
in Fig.~\ref{fig:smd}[B]) was obtained from  the integrated HI column density 
image at 26$''\times21''$ resolution  image and is well represented by a Gaussian:

\begin{equation}
\Sigma_{HI}(r)=\Sigma_0\times e^{-(r-c)^2/2r^2_0}
\label{eqn:hisb1}
\end{equation}

with $r_0=34.2^{''}\pm0.7^{''}$, c=$19.2^{''}\pm0.8^{''}$
and $\Sigma_0=7.8\pm0.1 \rm{M_\odot}~pc^{-2}$.

\subsection{HI Kinematics}
\label{ssec:HI_Kin}

\begin{figure}
\epsfig{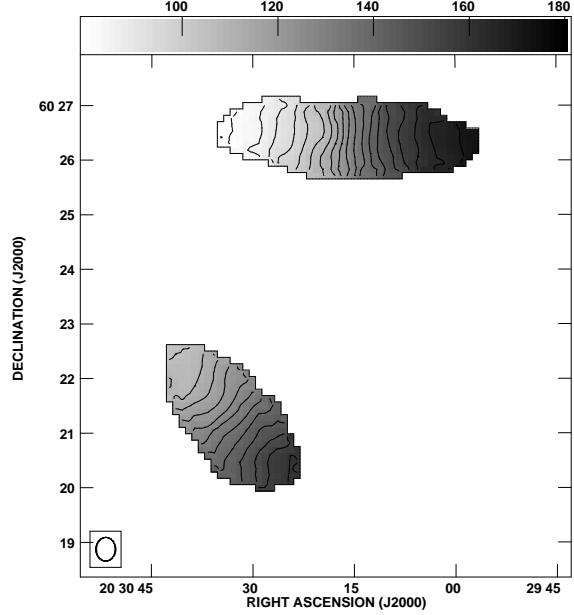}
\caption{The HI velocity fields of KK98 250 and KK98 251  at 26$^{''}\times 21^{''}$ 
          resolution. The contours are in steps of 5~\kms and
          range from 80.0~\kms to 170.0~\kms. 
        }
\label{fig:mom1}
\end{figure}

The velocity fields of KK98 250 and KK98 251, derived from the 26$''\times21''$ resolution
HI data cube are shown in Fig.~\ref{fig:mom1}. The velocity fields are regular and 
the isovelocity contours are approximately parallel, which is a signature of rigid 
body rotation.

     The velocity field of KK98 251 is slightly lopsided, the isovelocity  contours 
in  the north-eastern half of the galaxy are more curved than the south-western 
half. Some small scale kinks are also seen in the isovelocity contours in the outer
regions of the galaxy, indicative of a weak warping of the outer disk. These  
effects are more prominent  in the higher spatial resolution velocity fields 
(not shown), which also show a mild twist in the kinematical major axis in the 
outer region of the galaxy, consistent with warping.

\subsection{HI Rotation Curve}
\label{ssec:rotcur}

The rotation curves of the galaxies were derived separately for each of the
$26^{''}\times21^{''}$, $16^{''}\times14^{''}$ and $11.5^{''}\times10^{''}$
velocity fields, using the usual tilted ring model (Warner et al. 1973).
For the edge on galaxy KK98 250, as described in more detail below, the rotation 
curve was also derived by fitting to the position-velocity diagram.

\begin{figure}[]
\epsfig{file=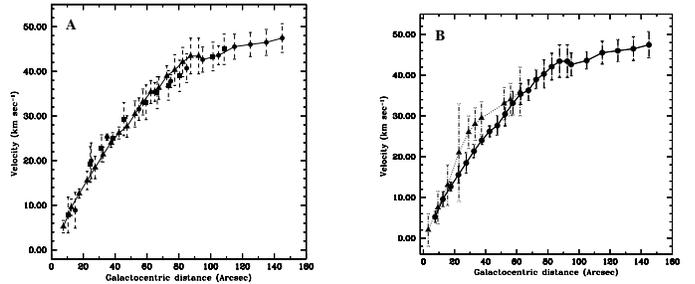,width=3.6in}
\caption{
{\bf [A]} The rotation curves for KK98 250 derived from the intensity weighted velocity 
     field at various resolutions. Circles, squares and triangles represent the rotation
     velocity derived from the $26^{''}\times21^{''}$, $16^{''}\times14^{''}$ 
     and $11.5^{''}\times10^{''}$ resolution  respectively. The adopted rotation 
     curve is shown by solid line. 
{\bf [B]} The adopted HI rotation curve for KK98 250 (dots) along with the H$\alpha$ rotation 
          rotation curve derived by~\cite{deblok01} (triangles). 
}
\label{fig:vrot1}
\end{figure}

      For each galaxy, the center and systemic velocity obtained from a global fit to 
the various resolution velocity fields matched within the error bars; the systemic 
velocity also matched  with the values obtained from the global HI profiles. Keeping the
center and systemic velocity fixed, we fitted for the inclination and position
angle (PA) in each ring. For KK98 250, keeping inclination as a free parameter in
the tilted ring fit gave unphysical results, hence, the kinematical inclination
of the galaxy was fixed to the  value estimated from the HI morphology, 
viz $80^\circ$. Note that at such high inclination angles, the uncertainty in 
inclination has only a small effect on the derived rotation curve (Begeman 1989). 
For example, if we fix the inclination to $70^\circ$, the derived rotation curve 
is the same, within the error bars. Fits for the position angle gave a value 
of $\sim 267^\circ$ (in good agreement with that derived from the optical image),
with no systematic variation across the galaxy. The rotation curves (derived 
with the PA and inclination fixed to the values of $267^\circ$ and $80^\circ$) are
shown in Fig.~\ref{fig:vrot1}[A]. Note that the rotation curves derived 
from the different resolution velocity fields match within the errorbars,
suggesting that, in spite of being highly inclined,  the effects of beam 
smearing are not significant. The  final adopted rotation curve is shown in 
Fig.~\ref{fig:vrot1}[A] as a solid line.  

\begin{figure}[]
\epsfig{file=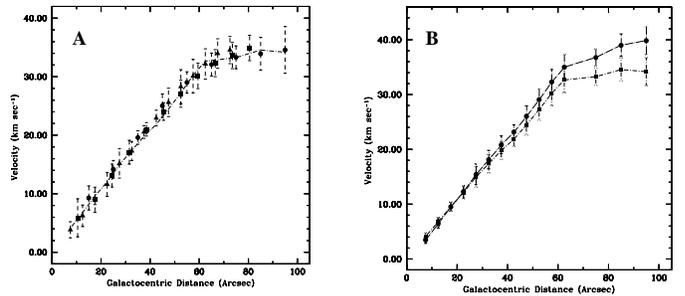,width=3.6in}
\caption{ 
{\bf[A]} The rotation curves for KK98 251 derived from the intensity weighted 
        velocity field at various resolutions. Circles, squares  and triangles
         show the rotation velocity derived from the $26^{''}\times21^{''}$, 
         $16^{''}\times14^{''}$ and $11.5^{''}\times10^{''}$ resolution  
         respectively. The adopted mean rotation curve is shown by dash-dot
         line. 
{\bf[B]} The adopted mean rotation curve for KK98 251 (squares) and  the 
         rotation curve after applying ``asymmetric drift" correction 
         (circles). See the text for more details.
}
\label{fig:vrot2}
\end{figure}

       For KK98 251, the inclination was found to be  $65^{\circ}$ (which agrees with 
that derived from ellipse fitting to the HI morphology, see Sect.~\ref{ssec:HI_dis}),
with  no systematic variation with the radius. The best fit position angle was
$\sim 230^\circ$ at all radii, except in the outermost regions of the galaxy where 
it changes to $\sim 215^\circ$. Ignoring this relatively small change, (which in any
case has negligible effect on the derived velocities),  
the rotation curve was derived by keeping the inclination and PA fixed at $65^{\circ}$ 
and $230^{\circ}$ respectively. Fig.~\ref{fig:vrot2}[A] shows the rotation curves for  KK98 251, 
derived from the various resolution velocity fields. Note that all the rotation curves 
match within the errorbars. Recall that KK98 251 shows slight kinematical lopsidedness.
Consistent with this, the rotation curve derived independently for the approaching
and receding sides are slightly different, although the difference is everywhere less
than 3.5~\kms. For the purpose of mass modeling, we have used a mean of
the rotation curves for the two sides. The errorbars on the mean rotation 
curve were obtained by adding  quadratically  the uncertainty reported by  the 
tilted-ring fit as well as the difference in rotation velocities between the 
approaching and receding side.

\begin{figure}[]
\epsfig{file=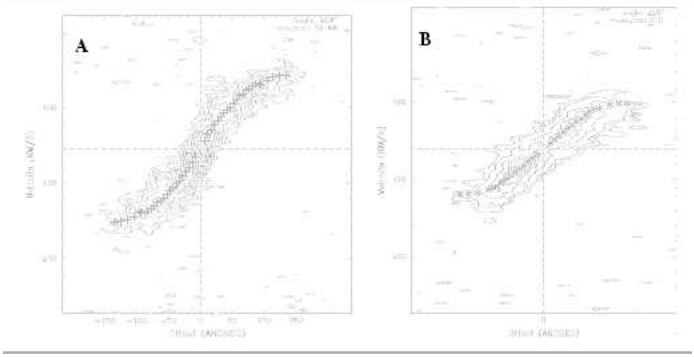,width=3.6in}
\caption{
  {\bf[A]} The  PV diagram of the galaxy along the kinematical
        major axis for KK98 250, with the adopted rotation curve overlayed.
  {\bf[B]} The  PV diagram of the galaxy along the kinematical
        major axis for KK98 251, with the adopted rotation curve overlayed.
}
\label{fig:PV}
\end{figure}

\begin{figure}[]
\rotatebox{-180}{\epsfig{file=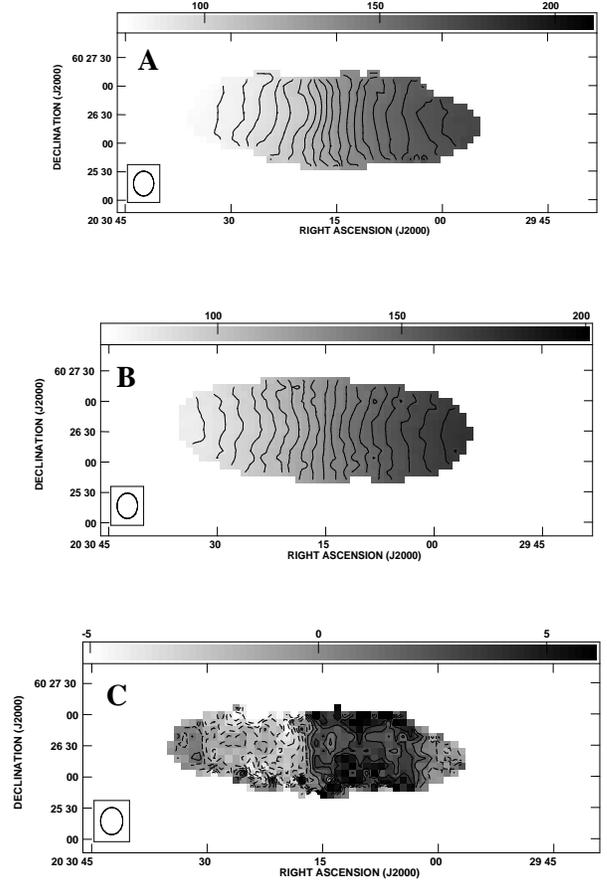,width=3.7in}}
\caption{ \textbf{[A]}The observed  velocity field of KK98 250 at 
         $26^{''}\times21^{''}$ resolution. \textbf{[B]}The model 
         velocity field derived from the rotation curve at 26$''\times21''$ 
         resolution. The contours are in steps of 5~\kms and range 
   from 81.0~\kms to 171.0~\kms.\textbf{[C]} The residual velocity field ([A]-[B]).
   The contours are in steps of 1.0~\kms and range from -4.0~\kms to 4.0~\kms.
  }
\label{fig:model1}
\end{figure}

      For highly inclined galaxies, rotation velocities derived from the titled ring
fits to the velocity field could underestimate the true rotation velocities, hence 
in such cases the  rotation curve is often estimated by fitting to the high velocity
edge of the emission  (e.g. Sancisi \& Allen 1979). While this method is well suited 
to large galaxies with flat rotation curves, it is not appropriate for
galaxies with solid body rotation.  Both the velocity field of KK98 250 as well as the
rotation curve derived from the tilted ring fit indicate solid body rotation,
hence, instead of using the `edge fitting' technique, the rotation curve for KK98 250 
was derived by interactively fitting to the PV diagram, using the task INSPECTOR in GIPSY 
(see Fig~\ref{fig:PV}[A]). The PA and the inclination in the interactive fitting 
were fixed to the values used in the tilted ring fit. The derived rotation curve 
matches, within the errorbars, to that derived from the tilted ring fit. As a 
further check of the robustness of the derived rotation 
curve, a model data cube for KK98 250 was constructed  using the  adopted rotation
curve and the observed HI column density profile, with the task GALMOD in GIPSY.
The model data cube was smoothed to a beam of 26$''\times21''$ resolution  
and the moment maps were derived in the same manner as for the real data. 
Fig.~\ref{fig:model1}[B] shows the derived model velocity field for KK98 250.
A residual (data-model) velocity field is shown in Fig.~\ref{fig:model1}[C];
as can be seen, the model velocity field provides a good match to 
the observed field.

Although KK98 251 is less inclined, a similar exercise of estimating the 
rotation velocities  from the PV diagram was repeated for it. 
Fig~\ref{fig:PV}[B] shows the adopted mean rotation curve projected 
onto the PV diagram -- as can be seen, the mean rotation curve provides 
a reasonably good fit to the data. A model data cube for KK98 251 was also 
constructed using the derived rotation curve, in the same manner as for 
KK98 250. Again, a good match between the model (not shown) and the 
observed field was found.

        The sensitivity of HI observations to the inner slope of the rotation
curve has been the subject of much recent discussion (e.g. van den Bosch 
\& Swaters, 2001). One possible way of overcoming the relatively poor 
resolution offered by HI observations is to instead use H$\alpha$ observations.
For KK98 250, an H$\alpha$ based rotation curve has been 
derived by ~\cite{deblok01}, and is shown with triangles 
in Fig.~\ref{fig:vrot1}[B]. As can be seen, although the H$\alpha$ curve 
is steeper than the HI curve at intermediate radii, in 
the innermost regions of the galaxy, (where the effects of beam smearing 
are expected to be most severe), the two rotation curves in fact 
show an excellent agreement.

As mentioned earlier, if one wishes to use the rotation curves to estimate the
total dynamical mass, then one needs to account for the pressure support of the
HI disk; this correction (generally called  the ``asymmetric drift'' correction) 
is given by:

\begin{equation}
\rm{ v^2_c=v^2_o - r}\times{\sigma}^2\bigl[\frac{\rm d}{\rm dr}(\ln{\Sigma_{\rm{HI}}})+\frac{\rm d}{\rm dr}(\ln{\sigma}^2)-\frac{\rm d}{\rm dr}(\ln{2\rm{h_z}})\bigr],
\label{eqn:ad}
\end{equation}

where $\rm{v_c}$ is the corrected circular velocity, $\rm{v_o}$ is the observed rotation velocity,
$\sigma$ is the velocity dispersion, and $\rm{h_z}$ is the scale height of the disk. Strictly 
speaking, ``asymmetric drift" corrections are applicable to collisionless stellar systems 
for which the magnitude of the random motions is much smaller than that of the rotation
velocity. However, it is often used even for gaseous disks, where the assumption
being made is that the pressure support can be approximated as the gas density times the
square of the random velocity. The observed profile width can be used as an estimator
of the velocity dispersion, after correcting for instrumental broadening and the finite
size of the synthesized beam. This procedure doesn't work for very inclined galaxies,
so, as discussed in Sect.~\ref{sec:obs} for KK98 250 we assume a constant velocity 
dispersion of 8~\kms. For KK98 251 we find that,  after putting in these corrections, the
estimated velocity  dispersion is also $\approx 8$~\kms. Further, in the absence of 
any measurement for $\rm{h_z}$, we assumed $\rm{d}(\ln(\rm{h_z}))/\rm{dr}=0$ (i.e. that the scale height 
does not change with radius). Substituting these values back in the Eqn.~(\ref{eqn:ad})
and using the fitted Gaussian profile to the radial surface density distribution, 
the ``asymmetric drift" correction was calculated and applied to the observed 
rotation velocities. For KK98 250 this correction is found to be small (less than 2.5~\kms),
compared to the errorbars on the rotation curve. On the other hand, the correction 
for KK98 251 is significant in the outer regions.  The ``asymmetric drift" corrected 
curve for KK98 251 is shown in Fig.~\ref{fig:vrot2}[B]. 

\subsection{Mass Model}
\label{ssec:massmodel}

\begin{figure*}[t!]
\rotatebox{-90}{\epsfig{file=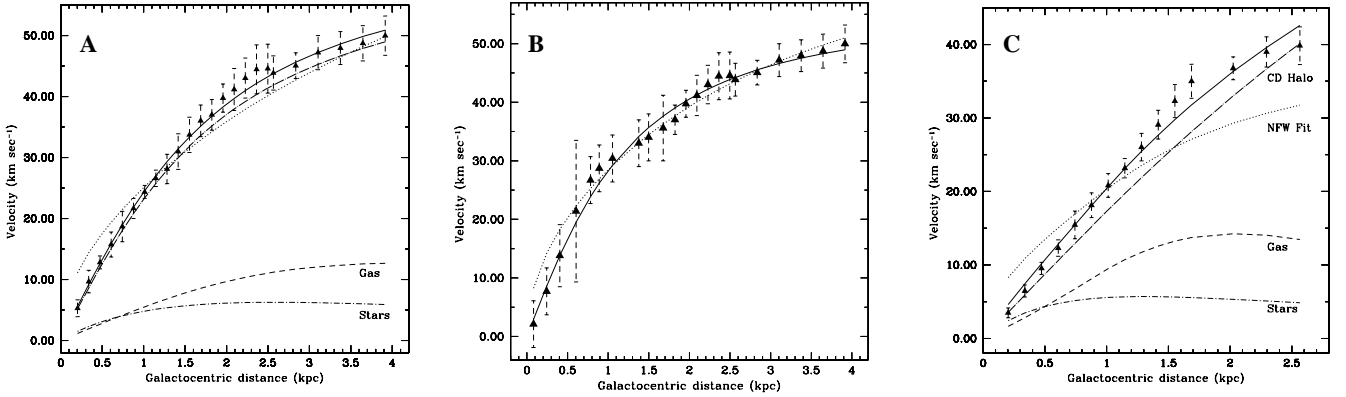,width=2.3in}}
\caption{{\bf[A]}Mass models for KK98 250 using the corrected rotation curve.
         The points are the observed data. The total mass of gaseous disk (dashed line)
          is $3.6\times10^6 \rm{M_\odot}$.The stellar disk (short dash dot line) has
          $\Upsilon_V=0.2$, giving a stellar mass of $3.4 \times10^7 \rm{M_\odot}$. The
          best fit total rotation curve for the constant density halo model is shown as
          a solid line, while the contribution of the halo itself is shown as a
          long dash dot line (the halo density is $\rho_0=37.6\pm1.2~\times10^{-3}
          ~\rm{M_\odot}$~pc$^{-3}$ and core radius r$_c=1.5\pm0.2$). The best fit 
          total rotation curve for an NFW type halo, using $\Upsilon_V=0.0$, 
          c=1.0 and $v_{200}$=331.0~\kms is shown as a dotted line. 
         {\bf[B]}The best fit mass model for an isothermal halo (solid line) 
 to the hybrid rotation curve of KK98 250 consisting of the H$\alpha$ rotation velocities in
the center and the ``asymmetric drift" corrected HI data in the outer regions
of the galaxy. Also shown in the figure is the minimum disk fit mass model for an
NFW type halo (dotted line). {\bf[C]}Mass models for KK98 251 using the corrected rotation curve.
     The total mass of gaseous disk 
          is $9.75\times10^7 \rm{M_\odot}$. The stellar disk  has
          $\Upsilon_I=0.5$, giving a stellar mass of $1.2\times10^7 \rm{M_\odot}$. The
          best fit  constant density halo model is $\rho_0=15.7\pm1.3~\times10^{-3}~\rm{M_\odot}$~pc$^{-3}$ 
and core radius r$_c=4.2\pm1.6$ (solid line). Also shown is the best fit total rotation 
     curve for an NFW type halo, using $\Upsilon_I=0.0$, c=1.0 and $v_{200}$=180.0~\kms as
     a dotted line. See the text for more details.
}
\label{fig:massmodel2}
\end{figure*}

In this section, we use the ``asymmetric drift" corrected rotation curves,
derived in the last section, to derive mass models for KK98 250 and KK98 251.  

         The contribution of the stellar mass to the observed rotation curves 
were  computed by assuming the galaxies  to have  exponential stellar disks, with 
a constant mass to light ratio ($\Upsilon$) and an intrinsic thickness ratio 
(q$_0$) of 0.2. We further assumed that the density distribution in the vertical 
($z$) direction falls off like sech$^2(z/z_0)$, with $z_0$ independent of 
galacto-centric radius (see e.g. van~der~Kruit \& Searle 1981, de~Grijs \& Peletier 1997). 
The contribution of the gaseous disks to the observed rotation curves were 
calculated using the observed HI surface mass density  profiles, with the HI surface
density being scaled by a factor of 1.4 to account for the contribution from
helium. There is little evidence that dwarf galaxies contain substantial amounts 
of molecular gas (e.g.~\cite{israel95},~\cite{taylor98}), hence, no correction 
was made for molecular gas. We also neglected the contribution of ionized gas, 
if any. Since there is some evidence for similar vertical distributions of the HI 
and stellar disks (e.g. Bottema et al. 1986), we assumed  the  HI disks also to
have a sech$^2(z/z_0)$ vertical profile, with an intrinsic thickness ratio of 
$q_0=0.2$. The circular velocities of the disk components were computed using the 
formulae given by~\cite{casertano83}. For the dark matter halo, we considered two 
types of density profiles, viz. the modified isothermal profile and the NFW 
(Navarro et al. 1996) profile.  For mass models using a modified isothermal halo,
the free parameters are the halo central density $\rho_0$, core radius r$_c$
and the mass to light ratio of the  stellar disk, $\Upsilon$. For the NFW models,
the free parameters are the halo concentration parameter c, $v_{200}$ (the circular velocity
at the radius at which the halo density is 200 times the critical density) and
the mass to light ratio of the stellar disk, $\Upsilon$.  Mass model were fit using the 
GIPSY task ROTMAS. 

\begin{table*}
\caption{Mass decomposition using isothermal halo}
\label{tab:halo}
\begin{tabular}{|l|ccc|cccc|}
\hline
 && {\bf{KK98 250}}    &&&& {\bf{KK98 251}}  &\\
{\bf{HI Curve}}  & $\Upsilon_V$&r$_c$ (kpc)&$\rho_0~(10^{-3}~\rm{M}_\odot$~pc$^{-3})$& &$\Upsilon_I$&r$_c$ (kpc)&$\rho_0~(10^{-3}~\rm{M}_\odot$~pc$^{-3})$\\\hline
\hline
Best fit  & 0.20$\pm$0.05&$1.5\pm0.2$&$37.6\pm1.2$&&$-$&$-$&$-$ \\
Maximum disk  & 3.0&$4.1\pm1.0$&$10.0\pm1.1$&&1.0&$10.5\pm14.5$&$13.4\pm1.6$ \\
From observed $<$V-I$>$  & 0.7&$1.6\pm0.1$&$30.4\pm1.0$ && 0.5&$4.2\pm1.6$&$15.7\pm1.3$ \\
Minimum disk &0.0&$1.4\pm0.04$&$39.0\pm1.0$& & 0.0&$2.8\pm0.5$&$18.5\pm1.0$ \\
Hybrid (HI+H$\alpha$) Curve &0.2 &$0.9\pm0.1$&$65.4\pm5.6$&&$-$ &$-$&$-$\\
H$\alpha$ Curve &0.0 &$0.63\pm0.08$&$117.8\pm16.5$&&$-$ &$-$&$-$\\
\hline
\end{tabular}
\end{table*}

\subsection{KK98 250}
        Since we could trace emission only from the brightest central regions of the
galaxy in the I band, we use the more accurately determined V band scale length for the
mass modeling. Fig.~\ref{fig:massmodel2}[A] shows the best fit mass models for KK98 250. The 
derived halo parameters are given in Table~\ref{tab:halo}. For comparison, apart from
the best fit mass model, the derived halo parameters are also given for the maximum disk,
minimum disk and $\Upsilon_V=0.7$ (which was obtained from the observed color 
$<$V-I$>$ of $\sim 1.3$ using the low metallicity Bruzual $\&$ Charlot SPS model using 
a modified Salpeter IMF; Bell $\&$ de Jong 2001). The total dynamical mass of KK98 250, 
estimated from the last measured point of the rotation curve is 
$22.6\times 10^8 \rm{M}_\odot$ --at this radius more than 92\% of the 
mass  of KK98 250 is dark. 

For mass models with an NFW halo, keeping $\Upsilon_V$ as a free parameter 
in the fit gave unphysical results. Even after setting $\Upsilon_V=0$, no reasonable 
fit could be obtained. As an illustration, Fig.~\ref{fig:massmodel2}[A] shows an NFW fit 
to the data, keeping the concentration parameters c fixed to 1, $\Upsilon_V$=0.0 and 
$v_{200}$ chosen to minimize $\chi^2$. As can be seen, even at these extreme values
for the parameters, the quality of fit is poor. 

    We also fit mass models to a hybrid rotation curve (see Fig.~\ref{fig:vrot1}[B]), 
consisting of H$\alpha$ data derived by~\cite{deblok01} in the inner regions of the 
galaxy and the ``asymmetric drift" corrected HI rotation curve in the outer regions. 
Again, keeping $\Upsilon_V$ as a free parameter in the fit gave unphysical results, hence
it was fixed to the value of 0.2, obtained from the best fit isothermal halo model, 
derived using the HI rotation curve alone. In any case, fixing $\Upsilon_V$ to a
common value allows a meaningful comparison of the halo parameters derived using 
both the rotation curves. The derived halo parameters for the isothermal halo 
are given in Table~\ref{tab:halo}. The table also shows the isothermal halo 
parameters derived by~\cite{deblok01} using only the H$\alpha$ rotation curve. 
We note that apart from the (probably not physically meaningful) maximum disk
case, the halo parameters derived from the HI rotation curve are in good
agreement with one another, but that they are substantially different from 
the parameters derived from the Hybrid or H$\alpha$ rotation curves. In this 
context it is worth repeating, that the discrepancy between the H$\alpha$ 
rotation curve and the HI rotation curve is largest at intermediate radii, and 
not at small radii as one would have expected, if the HI rotation curve 
suffered from beam smearing.

Also shown in Fig.~\ref{fig:massmodel2}[B] is the $\Upsilon_V=0$, NFW halo fit
to the Hybrid rotation curve. As can be seen, this fit overestimates
the observed rotation velocity in the inner regions of the galaxy; fixing 
$\Upsilon_V$ to more reasonable values worsens this discrepancy.
The NFW halo parameters (corresponding to $\Upsilon_V=0$) are c=6.5$\pm1.4$ and 
$v_{200}=66.0\pm13.0$. The value of $6.5$ for the concentration parameter is
much lower than the range of NFW concentration parameters predicted by LCDM 
simulations (Bullock et al. 2001). It is worth mentioning here that even for
the H$\alpha$ rotation curve alone, isothermal halos provide a substantially
better fit than an NFW halo (de~Blok et al. 2001).

\subsection{KK98 251}

As discussed in section~\ref{ssec:opt_obs} for KK98 251, we use the I band scale length
derived by Karachentsev et al. (2000). For isothermal halo models, keeping 
$\Upsilon_I$ as a free parameter in the fit gave negative values for $\Upsilon_I$. 
Further, if one keeps $\Upsilon_I$ fixed, the $\chi^2$ continuously decreases 
as $\Upsilon_I$ is decreased.  Fig.~\ref{fig:massmodel2}[C] shows the best fit 
mass model for KK98 251 for a constant density halo using $\Upsilon_I=0.5$ (which
corresponds to the observed $<$V-I$>$ color, from  the low metallicity Bruzual $\&$ 
Charlot SPS model using a modified Salpeter IMF). The derived halo parameters
using various values of $\Upsilon_I$ are given in Table~\ref{tab:halo}. As can 
be seen from Table~\ref{tab:halo}, the halo parameters are relatively 
insensitive to the assumed value of $\Upsilon_I$. The total dynamical mass (at
the last measured point of the rotation curve) is $\rm{M_T}$=9.5$\times 10^8 \rm{M_\odot}$.
For KK98 251, we found that even for $\Upsilon_I=0$, with no value of c and $v_{200}$ 
could a good fit using an NFW halo be obtained. The best fit NFW model is also
shown in Fig.~\ref{fig:massmodel2}[C]; as can be seen, the data deviate substantially
from the model.

\begin{figure}[]
\rotatebox{-90}{\epsfig{file=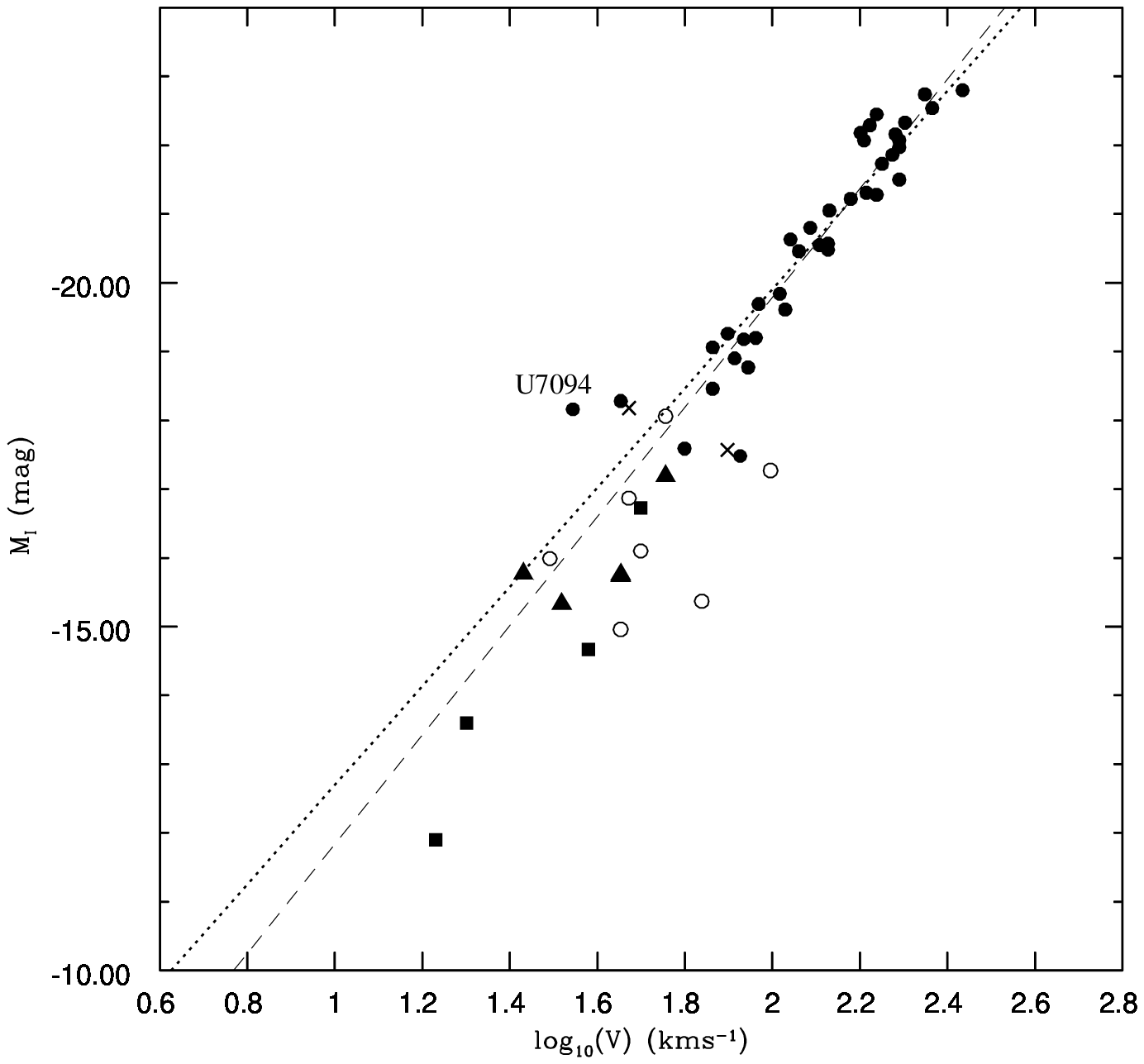,width=2.6in}}
\caption{  I magnitude Tully-Fisher diagram. The filled circles are from~\cite{verheijen01}, 
    empty circles from~\cite{swaters99} with I band magnitude from~\cite{patterson}, 
    filled triangles  from~\cite{swaters99} with I band magnitude from~\cite{makarova98} and 
    crosses from~\cite{deblok96} with I band from~\cite{patterson}. Filled squares are from
    galaxies in our sample including KK98 250, KK98 251, CamB and DDO210. The dotted line is 
    the best fit to~\cite{verheijen01} sample, while the dashed line is the best fit
    to the ~\cite{verheijen01} sample excluding the galaxy U7094. See the text for more details.
  }
\label{fig:TF_relation}
\end{figure}

\subsection{The Tully-Fisher relation}
\label{ssec:discuss}

          It has been recently suggested that dwarf galaxies deviate systematically 
from the Tully-Fisher (TF) relation defined by bright galaxies (i.e.~\cite{stil99},
\cite{swaters99},~\cite{mcgaugh00}), with small galaxies being underluminous 
compared to what would be expected had they followed the same TF relation as 
$\sim L_*$ galaxies. Contrary to this suggestion, ~\cite{pierini99} found no 
evidence of any break in the  near-IR TF relation  for faint dwarf galaxies. 
The reason for this discrepancy is unclear. We note however that high 
resolution HI images are probably crucial in studies of the TF relation for 
very faint galaxies. This is because (i)~the inclination may often be difficult 
to obtain from images of faint irregular galaxies, and (ii)~the 50\% HI velocity 
width may not be a good indicator of the rotation velocity in faint dwarf galaxies,
where random motions are comparable to the peak rotational velocities 
(e.g. Camelopardalis B;~\cite{begum03}, DDO210;~\cite{begum04}). 
For such galaxies, it is important to accurately correct for the pressure 
support (``asymmetric  drift'' correction), which is possible only if one 
has high resolution HI images. Unfortunately, there are only few dwarf 
galaxies with both I band photometry and HI synthesis imaging available. 
Interestingly enough, even without this correction, (or even a correction 
for inclination), the HI velocity widths of dwarf galaxies with well
measured distances do correlate with the absolute blue luminosity, 
albeit with large scatter (Huchtmeier et al. 2003). It is currently a 
matter of speculation as to whether the scatter would reduce 
significantly, once one puts in the corrections discussed above.

  We show in Fig.~\ref{fig:TF_relation}, the TF relation for a sample of galaxies 
(with both HI synthesis imaging and I band photometry) spanning a range of 
I magnitude from $\sim-$23.0 to $-$12.0.  For the bright galaxies, the photometry 
and rotation velocities are from ~\cite{verheijen01}, while for the fainter galaxies,
the I band magnitudes are from Patterson \& Thuan (1996) and  
Makarova \& Karachentsev (1998), while the rotation velocities were taken 
from Swaters (1999) and de Blok et al. (1996). The faintest two galaxies 
(Camelopardalis~B and DDO210) are from our 
earlier work (Begum et al. 2003;  Begum \& Chengalur 2004).  For Camelopardalis~B,
the I magnitude is calculated using the relation (B$-$V)$_{\rm{T}}=0.85$(V$-$I)$_{\rm{T}}-$0.2 
(Makarova \& Karachentsev 1998).  As can be seen from Fig.~\ref{fig:TF_relation},
faint dwarf galaxies do tend to lie systematically below the  TF relation defined 
by bright galaxies (dotted line). This trend persists even if one excludes the 
most discrepant bright galaxy (U7094, which has a poor quality rotation curve)
in determining the bright galaxy TF relation (dashed curve). Note also that
the scatter in the TF relation increases at the faint end -- this may be in 
part due to the uncertainty in the distances to the galaxies.

\section{Conclusions}

To conclude, we have presented  high velocity resolution  HI 21 cm 
synthesis images of the dwarf galaxies, KK98 250 and KK98 251, as well as 
optical broad band images of KK98 250. We find that the HI disks of the 
galaxies do not show any clear signs of tidal disturbance and that 
both galaxies have regular velocity fields, consistent with rigid body rotation.
We fit the rotation curves with dark matter halo mass models and find 
that both rotation curves can be fit using modified isothermal halos
but not with NFW halos. Finally, for a small sample of galaxies with 
both I band photometry and HI synthesis images, we find, in agreement 
with earlier studies (which used single dish HI data), 
that dwarf galaxies tend to lie below the I band TF relation defined 
by brighter galaxies.

\begin{acknowledgements}

        The HI observations presented in this paper would not have been possible without 
the many years of dedicated effort put in by the GMRT staff in order to build the 
telescope. The GMRT is operated by the National Centre for Radio Astrophysics of 
the Tata Institute of Fundamental Research.  We thank the staff of IAO, Hanle 
and CREST, Hosakote, that made the optical observations possible. The facilities at IAO 
and CREST are operated by the Indian Institute of Astrophysics, Bangalore. We are 
also grateful to Dr Devendra Kumar Sahu invaluable help with the optical 
observations.

\end{acknowledgements}

\end{document}